# Field tunable BKT and quantum phase transitions in spin-$\frac{1}{2}$ triangular lattice antiferromagnet

Dechen Zhang[1], Yuan Zhu[1], Guoxin Zheng[1], Kuan-Wen Chen[1], Qing Huang[2], Lingxiao Zhou[1], Yujie Liu[3], Kaila Jenkins[1], Aaron Chan[1], Haidong Zhou[2], and Lu Li[1*]

[1] *Department of Physics, University of Michigan, Ann Arbor, MI 48109, USA.*

[2] *Department of Physics, University of Tennessee, Knoxville, TN, USA*

[3] *Department of Materials Science and Engineering, University of Michigan, Ann Arbor, MI 48109, USA*

[*] Corresponding to: luli@umich.edu

Quantum magnetism is one of the most active fields for exploring exotic phases and phase transitions. The recently synthesized $Na_2BaCo(PO_4)_2$ (NBCP) is an ideal material incarnation of the spin-$\frac{1}{2}$ easy axis triangular lattice antiferromagnet (TLAF). Experimental evidence shows that NBCP hosts the spin supersolid state with a giant magnetocaloric effect. It was also proposed that the applied magnetic field *B* can drive the system through Berezinskii-Kosterlitz-Thouless (BKT) and other richer quantum phase transitions. However, the detection of these transitions is challenging because they onset at extremely low temperature *T* at around 60 mK, and the measurement of the magnetic susceptibility of these transitions requires high sensitivity. With the help of our newly developed gradient force magnetometer in a dilution refrigerator, we constructed the contour diagram of the magnetic susceptibility in the *B-T* phase diagram in *T* as cold as 30 mK. These results provide a more comprehensive and accurate understanding of the several field-tunable quantum phase transitions and BKT melting of the spin supersolidity, which are especially significant when their giant magnetocaloric effects highlight potential applications for sub-Kelvin refrigeration under concerns about global helium shortages.

Materials in which the electrons are strongly correlated display a wide variety of novel phenomena. Over the past decades, there has been considerable interest in frustrated quantum magnets[1-3]. A prominent example is the spin-$\frac{1}{2}$ triangular lattice Heisenberg antiferromagnet (TLHAF), which has been proven to be fertile ground for studying unconventional phase transitions. Due to the interplay of inherent geometric frustration and strong quantum fluctuation, the ground state property of TLHAF has always been a difficult and controversial problem, although intensively investigated[4,5]. Experimentally, ideal spin-$\frac{1}{2}$ TLHAF is rarely reported, such as $Cs_2CuBr_4$[6-10], $AYbCh_2$ (A = Na and Cs, Ch = O, S, Se) family[11-13], $Ba_3CoSb_2O_9$[14-17], and $Ba_3CoNb_2O_9$[18,19]. Although the characteristic up-up-down (UUD) state stabilized by quantum fluctuation has been observed in these compounds, none exhibit easy-axis anisotropy. It was not until recently that a cobalt-based compound $Na_2BaCo(PO_4)_2$ (NBCP) was synthesized[20], which was found to represent the first ideal material incarnation of the prototypical spin-$\frac{1}{2}$ easy-axis TLHAF XXZ model[21-24]. Moreover, the observation of supersolidity and its strong entropic effect at the phase transitions greatly enriched the research interest of this quantum magnetic system[25].

Supersolid, which consists of incompressible solid-structure particles while simultaneously showing superfluid order, has stimulated great research interest and activities[26-32]. Since $S = \frac{1}{2}$ spins can be mapped onto hard-core bosons, the spin supersolid state can be viewed as the quantum magnetic analogue of the supersolid state of Bose atoms that breaks both the lattice translational and spin $U(1)$ symmetries. Based on the general $S = \frac{1}{2}$ easy-axis TLHAF model[33], the prominent valley-like regimes hosting the spin supersolid states are proposed in the NBCP spin system[22,25]. In the $H$-$T$ phase diagram, NBCP successively goes through multiple magnetic phases with the application of an external magnetic field. These phases with broken and restored $U(1)$ symmetry are separated by several BKT transitions, which are beyond the Landau-Ginzburg paradigm of spontaneous symmetry breaking. Therefore, NBCP constitutes a rare and ideal quantum magnetic platform to investigate the BKT transitions in the three-dimensional spin system.

However, the relatively low saturation field of NBCP makes it a considerable challenge to detect these phase transitions, onsetting at extremely low $T$ at ~ 60 mK, as predicted by the simulation[22]. To track these transitions, we integrated the cantilever magnetometer into a dilution refrigerator with both applied magnetic field and magnetic field gradient. The so-called gradient force magnetometer can measure the exact value of the ground-state DC magnetization compared to traditional magnetic torque[9,34], down to the base temperature of the dilution refrigerator, with high sensitivity analogous to that of a commercial magnetometer. We verified that we could observe the 1/3 and saturation magnetization plateau, with the field gradient not influencing the ground state magnetic structure. We map the $B$-$T$ phase diagram of magnetic susceptibility down to 30 mK. We observed the magnetic phase transitions, including the BKT melting of the spin supersolidity and multiple field-tunable quantum phase transitions. These results offer an ideal material incarnation for investigating novel phase transitions of the general triangular lattice easy-axis frustrated magnet.

Fig. 1a illustrates the crystal structure of NBCP. Due to strong spin-orbital coupling, each magnetic $Co^{2+}$ ion possesses an effective spin doublet with $S = \frac{1}{2}$. These $Co^{2+}$ ions form a nearly perfect triangular lattice in the crystal's ab-plane under a slightly distorted octahedra crystal field environment. The primary spin exchange interactions between $Co^{2+}$ ions are governed by in-plane super-super-exchange. At the same time, the interlayer coupling is negligibly small due to the near-perfect cancellation of interactions between the nearest and next-nearest interlayer neighbors[23]. Consequently, NBCP can be microscopically characterized as an ideal $S = \frac{1}{2}$ TLHAF.

Previous research has investigated multiple low-temperature magnetic phases of NBCP under an out-of-plane magnetic field. As depicted in Fig. 1b, these phases include the UUD, Y, and V states. In the UUD state, two spins point up, and one spin points down within a magnetic unit cell, breaking the lattice translational symmetry while preserving $U(1)$ rotational symmetry. The Y and V states, characterized by spin configurations resembling the shapes of the letters Y and V, respectively, spontaneously break both the lattice translational symmetry—due to their in-plane spin components—and the $U(1)$ rotational symmetry. These states can also be interpreted using boson language through hardcore boson mapping, as shown in Fig. 1c. The UUD state is analogous to the gapped bosonic Mott insulator. The Y and V states correspond to supersolid phases that simultaneously show crystalline properties and superfluid orders.

The magnetic phases and phase transitions in NBCP are of theoretical interest and practical relevance for sub-Kelvin cooling applications. However, the complete phase diagram still needs to be solved, with discrepancies between experimental data and DMRG simulations. To further explore the low-temperature magnetic properties, we employed a gradient force magnetometer (GFM) to measure the DC magnetization of NBCP, as illustrated in the insert of Fig. 2a. A magnetic field $B$ and field gradient $\nabla B$ were applied along the $z$-direction. The NBCP sample, aligned with the crystalline $c$-axis or $a$-axis along the $z$-direction, was mounted on the cantilever's metal pad. Torque signals generated by the sample's intrinsic magnetization $M$ under $B$ and $\nabla B$ were detected capacitively using cantilevers with varying stiffness (see Methods).

Fig. 2a and b display the field dependence of capacitance measurements corresponding to the magnetization of NBCP along the crystalline $c$-axis and $a$-axis at selected temperatures. The raw data comprises two components: the $B$-antisymmetric gradient force signal $\tau_{\text{gradient}}$ and the $B$-symmetric torque signal $\tau_{\text{torque}}$. To extract the magnetization, we first calculate the total torque $\tau_{tot}$ and perform antisymmetrization to isolate $\tau_{\text{gradient}}(B)$, which is directly proportional to the sample magnetization $M$. The results are then normalized by the saturation magnetization $M_s$, yielding the isothermal magnetization curves in Fig. 1c ($B//c$) and d ($B//a$).

For the field along the $c$-axis (Fig. 2c), $M_c/M_s$ shows a plateau between 0.45 T ($B_{c1}$) and 1.07 T ($B_{c2}$). This plateau narrows as the temperature increases but persists up to 300 mK. Above 1.9 T ($B_{c3}$), all spins become fully polarized, and the magnetization saturates. These transition fields and temperatures align with prior AC susceptibility measurements. The observation of the magnetization plateau confirms that the GFM is an effective technique for measuring low-temperature magnetization without altering the magnetic ground state. However, additional features, such as a hysteresis loop and broad cusp in the down-sweep curves between $B_{c2}$ and 1.45 T, suggest the occurrence of quantum phase transitions. Fig. 2d presents the temperature dependence of $M_a/M_s$. In agreement with previous studies, NBCP is near isotropic in the $ab$-plane. At higher temperatures, the magnetization increases almost linearly with $B$ before saturating at 1.71 T ($B_{a2}$). Below ~220 mK, a hysteresis loop and a broad dip around 0.7 T in the down-sweep curves indicate the presence of phase transitions.

Given the rich and delicate magnetic phases in NBCP, both in magnetic field and temperature, we also measured the temperature dependence of $M$ using the GFM to capture all phase transitions and signatures at the boundaries. Fig. 3a shows the DC magnetic susceptibility $\chi_c$ of NBCP under various $B$ below 0.415 T along the c-axis. As the temperature decreases, $\chi_c$ exhibits two transitions across a broad range of $B$. The precise positions of these transitions were identified by taking the temperature derivative of $\chi_c$, as marked by dashed lines in Fig. 3b.

The first transition is observed as a peak in $d\chi_c/dT$, and onsets at lower temperature. As $B$ increases, the peak gradually weakens until it eventually vanishes. The peak position shifts to lower temperatures, delineating the boundary of the spin-supersolid $Y$ state. The second transition, known as the three-state Potts transition[35], appears as a dip in $d\chi_c/dT$ and marks the evolution from the low-temperature UUD ordered state to the high-temperature paramagnetic phase. The Potts transition temperature increases with $B$, shifting from ~ 0.15 K at $B = 0\,T$ to ~ 0.25 K at $B = 0.415\,T$, as shown in Fig. 3b. Below $B = 0.115\,T$, the dip becomes less and less pronounced near zero field (at $B = 0.065\,T$), suggesting that the Potts transition is obscured by another BKT transition with algebraic quasi-long-range order[36]. This transition is also evident in the prior specific heat data, where a broad peak near zero field corresponds to the onset of the BKT transition[21,22]. The BKT transitions are suppressed at higher fields, and a much sharper peak indicates the Potts transition.

Above 0.415 T, the sample was mounted on a stiffer cantilever, and the measured magnetization $M_c$ and its temperature derivative $dM_c/dT$ are shown in Fig. 4. c and d. As the magnetic field increases, the dip marking the Potts transition boundary of the UUD state shifts to higher temperatures, reaching a maximum value of 0.335 K at $B = 0.7\,T$. Between 0.7 T and 1.0 T, the temperature of the dip decreases with increasing field and forms a dome-shaped region. Beyond 1.1 T, the dip is replaced by a large bump in the $M_c$ curves between 1.1 and 1.3 T. This bump indicates the onset of a third phase transition from the UUD state to the spin-superfluid $V$ state. The position of this bump aligns with the hysteresis loop observed in Fig. 1c.

The multiple magnetic phases of NBCP are mapped in the $B$-$T$ phase diagram, as shown in Fig. 4a. At zero field, the system exhibits three BKT transitions, corresponding to two broad peaks previously observed in specific heat measurements. The lowest-temperature BKT transition occurs around 100 mK and marks the boundary between the UUD and spin-supersolid $Y$ states, which is consistent with the lower-temperature specific heat peak. Below this translation, the $Y$ state develops an in-plane spin component and spontaneously breaks the $U(1)$ symmetry. Above the transition, the UUD state restores the $U(1)$ symmetry. On the other hand, the UUD state exhibits six-fold anisotropy within the in-plane degree of freedom, attributed to its three sublattice structure. This six-fold anisotropy becomes insignificant at elevated temperatures, around the Potts transition at 150 mK, where a distinct $U^*(1)$ symmetry emerges in the system. The $U^*(1)$ symmetry leads to two BKT transitions similar to other TLHAFs: one breaking and the other restoring $U^*(1)$ symmetry[37-40]. Recent spectroscopic experiments also found evidence for Goldstone modes associated with the $U(1)$ symmetry and pseudo-Goldstone modes related to the $U^*(1)$ symmetry[41,42], providing support for these BKT transitions[43].

With an external magnetic field $B//c$, the phase diagram is characterized by multiple phase transitions, including a dome-shaped Potts transition and two valley-shaped BKT transitions. Between 0.15 K and 0.2 K, the Potts transition marks the change from the z-direction ordered UUD state to the paramagnetic phase, consistent with sharp specific heat peaks from earlier measurements[21]. At lower temperatures, two BKT transitions separate the UUD phase from two spin-supersolid phases. The UUD phase maintains $U(1)$ symmetry while breaking lattice translational symmetry. Through the BKT transitions, the system enters the $Y$ phase at lower fields (Fig. 4c) and the V phase at higher fields (Fig. 4d), breaking $U(1)$ and lattice translational symmetry. In boson language, these transitions are analogous to the evolution from a Bose Mott insulator state to a supersolid state.

When $B//a$, two distinct magnetic phases emerge as $B$ increases until the system becomes fully polarized. As shown in Fig. 1d, the critical field separating these two phases is $B_{a1} = 0.7$ T. The signature of this phase transition is also evident in the temperature dependence of $M_a$ in Fig. 5a. Below ~ 0.4 T, the $M_a$ curves exhibit a gradual variation with temperature and arrange evenly over $B$. Above 0.4 T, $M_a$ drops sharply as $T$ decreases below ~ 220 mK, with the drop being most pronounced between 0.7 and 0.9 T. This behavior corresponds to the large dip observed at $B_{a1}$ in the $B$-dependent data of $M_a$ shown in Fig. 1d. After the system enters the second phase above 0.9 T, the drop in $M_a$ becomes progressively smoother with increasing $B$. As $B$ reaches $B_{a2}$, the system transitions into a quasi-polarized phase.

The $B$-$T$ phase diagram is further plotted to clarify these three phases. According to DMRG simulations, these phases can be attributed to $\tilde{Y}$, $\tilde{V}$ and QP (quasi-polarized) states respectively. When $B//a$ is applied, it immediately breaks the $U(1)$ rotational symmetry with respect to the $c$ axis. As $B$ increases beyond $B_{a1}$, the system transitions from the $\tilde{Y}$ state to the $\tilde{V}$ state, breaking the $Z_2$ symmetry with respect to the x-axis. Both $\tilde{Y}$ and $\tilde{V}$ states are 6-fold degenerate, and the symmetrical incompatibility between the two states suggests a likely discontinuous, first-order transition. This is evidenced by the large hysteresis loop near $B_{a1}$, indicating the metastability associated with quantum phase transitions[44-47]. Additionally, this transition is prominent in both the $B$- and $T$-dependent data, a feature not seen in previous experiments.

To summarize, the $B$-$T$ phase diagram of NBCP presents a coherent picture of its complex magnetic phases, highlighting several spin-supersolid phases and delicate BKT and quantum phase transitions. The transitions demonstrate rich symmetry-breaking phenomena beyond the Landau-Ginzburg paradigm, which are crucial for understanding the fundamental spin-$\frac{1}{2}$ easy-axis TLHAF model. Furthermore, studying these phases is significant for entropy-based cooling technologies in the sub-Kelvin regime, especially amid global helium shortages. The advanced measurements with the GMF technique enable precise mapping of the magnetic phases and phase transitions, positioning NBCP as a unique platform for exploring novel quantum phase transitions and highly efficient low-temperature cooling technologies.

**Figure Caption**

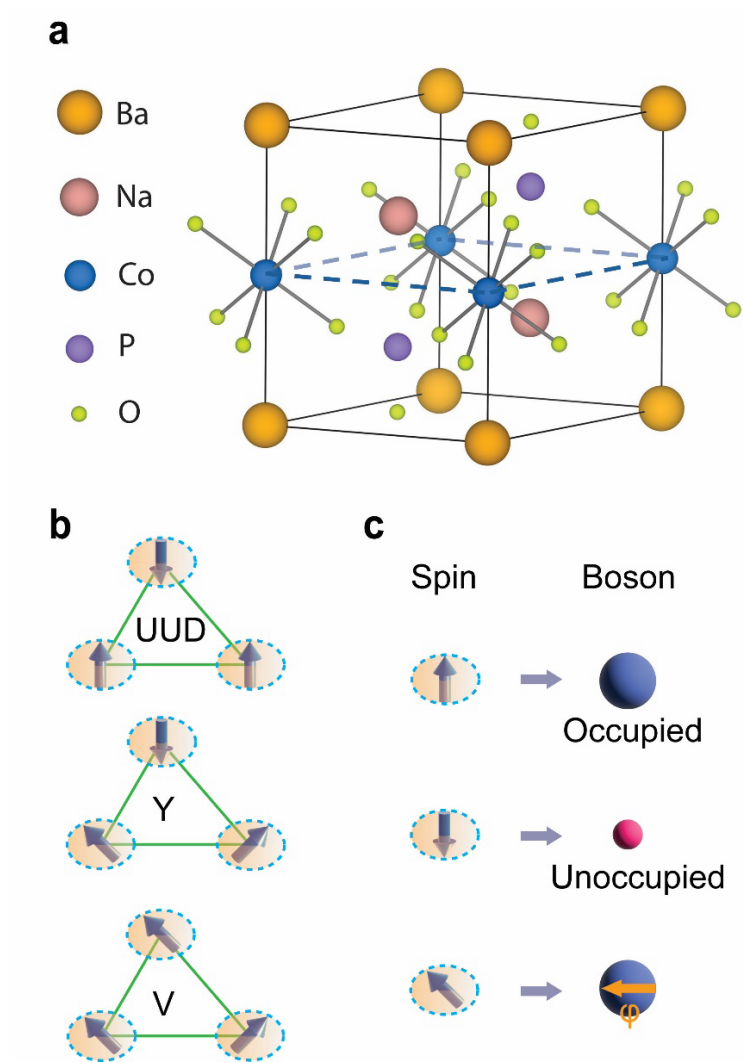

**Fig. 1 Illustration of crystal structure, spin configurations, and spin-boson mapping.** (a) Crystal structure of NBCP showing $Co^{2+}$ ions forming a triangular lattice in the *ab*-plane. (b) Multiple magnetic phases (UUD, *Y*, and *V* states) of NBCP, which are tunable under an out-of-plane magnetic field, showing different symmetry breakings. (c) The $S = \frac{1}{2}$ spins can be mapped into hardcore bosons. The spin-up and spin-down states (components) correspond to the occupied (blue ball) and unoccupied (red ball) boson sites. The in-plane spin components correspond to the additional phase $\varphi$.

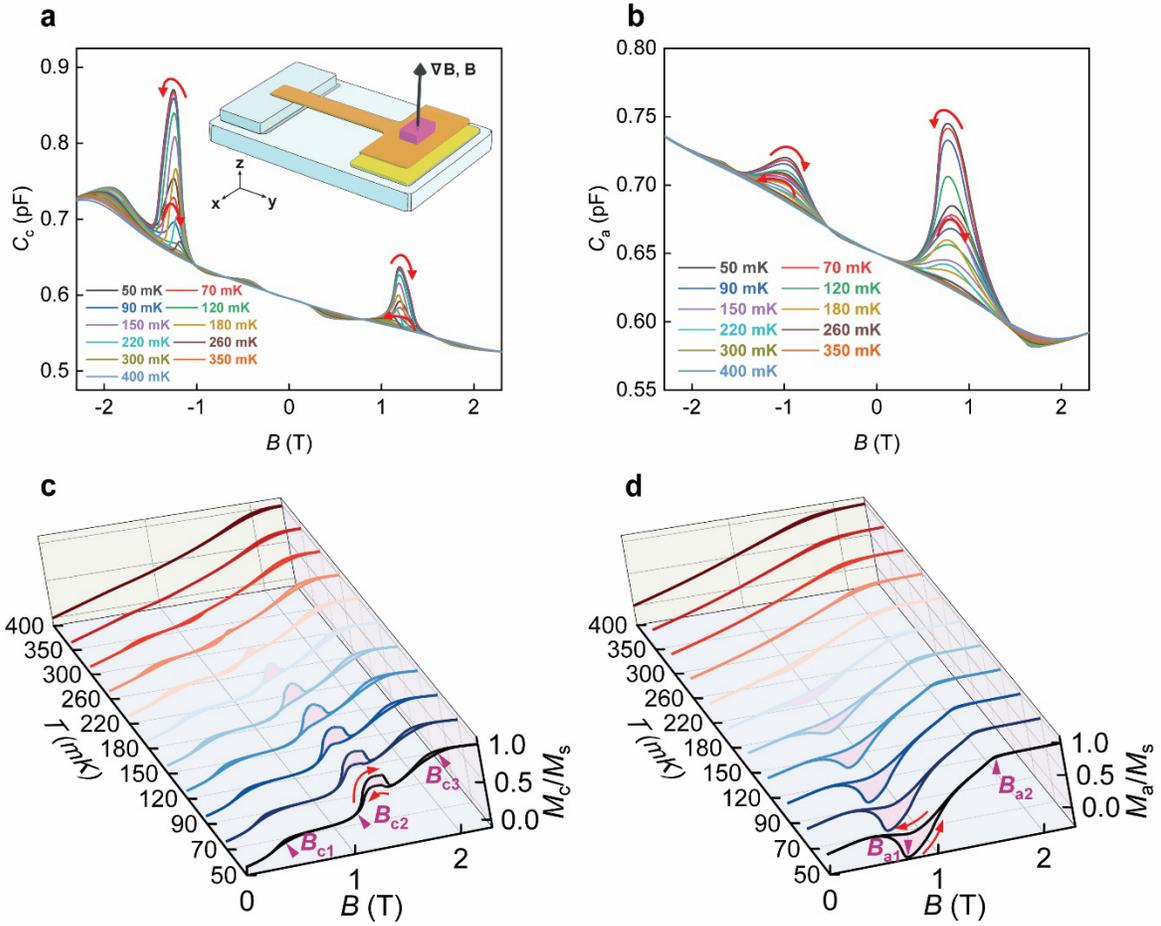

**Fig. 2 Experimental setup and *B*-dependent isothermal magnetization of NBCP.** (a) *B*-dependent capacitance readings with *B* along the c-axis at a fixed temperature. The inset depicts the gradient force magnetometer (GFM) setup using a cantilever and metal plate capacitor for DC magnetization measurements. (b) Capacitance measurements with *B* along the *a*-axis. (c) Isothermal magnetization along the *c*-axis after antisymmetrization and normalization shows a one-third magnetic plateau between 0.45 T and 1.07 T, persisting up to 300 mK. The pink region represents the hysteresis loop. (d) Isothermal magnetization along the *a*-axis. Below 220 mK, the dip around 0.7 T with hysteresis (pink region) indicates a phase transition.

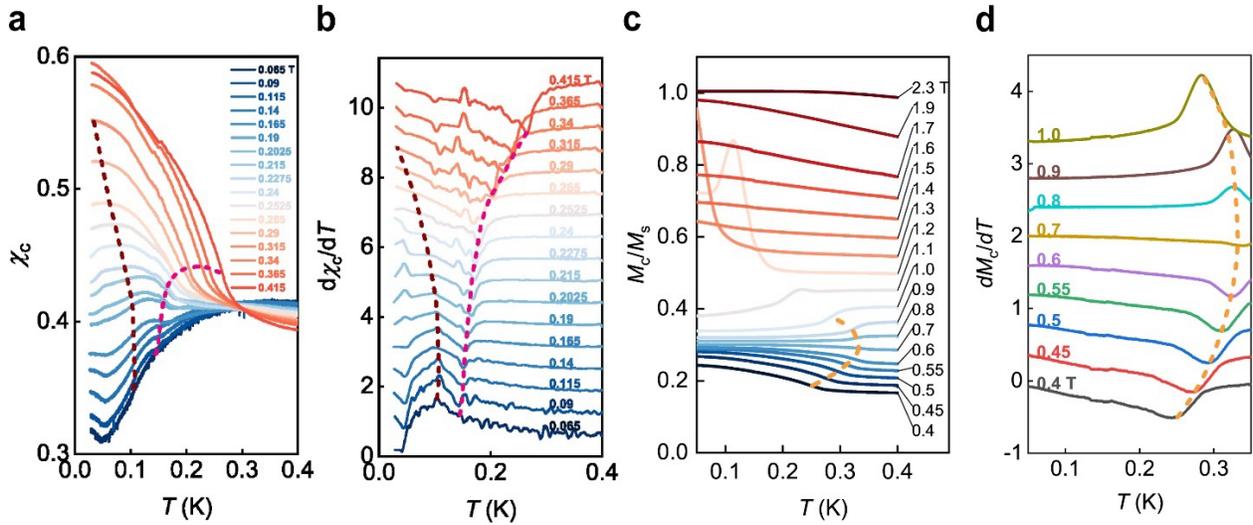

**Fig. 3 *T*-dependent DC magnetic susceptibility and magnetization and their derivative with respect to *T* along the *c*-axis.** (a) DC magnetic susceptibility $\chi_c$ of NBCP under various *B* below 0.415 T. The two transitions in $\chi_c$ are marked by the dashed line, indicating the Potts transition from the high-temperature paramagnetic state to the UUD state and the BKT transition from the UUD state to the spin-supersolid *Y* state. (b) Temperature derivative of $\chi_c$ ($d\chi_c/dT$). The two groups of peaks are used to identify the precise positions of the transitions. (c) The measured temperature dependence of the magnetization $M_c$ above 0.415 T with the sample mounted on a stiffer cantilever. (d) The temperature derivative $dM_c/dT$. As *B* increases, the dip marking the Potts transition boundary of the UUD state forms a dome-shaped region. Between 1.1 and 1.3 T, the bump in $M_c$ indicates the BKT transition from the UUD state to the spin-superfluid *V* state.

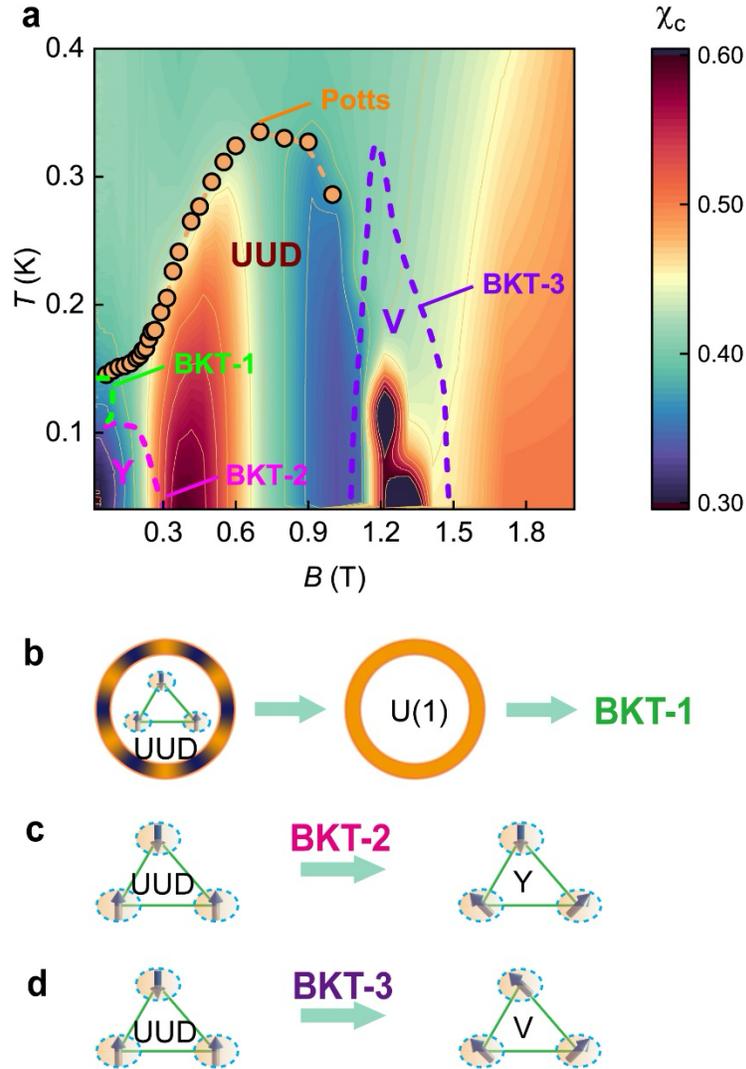

**Fig. 4 The *B-T* phase diagram and BKT transitions of NBCP with *B* along the *c*-axis.** (a) The phase diagram shows three BKT transitions and a Potts transition, separating the high-temperature paramagnetic, UUD, *Y*, and *V* states. (b) The two BKT transitions at zero field involve one transition that breaks $U(1)$ symmetry and the other that restores $U(1)$ symmetry. The occurrence of the BKT transitions is a consequence of the emergent $U(1)$ symmetry derived from the six-fold degenerate UUD state. (c) The low-field BKT transition from the spin-supersolid *Y* state (breaking $U(1)$ symmetry) to the UUD state (restoring $U(1)$ symmetry). (d) The high-field BKT transition from the spin-supersolid *V* (breaking $U(1)$ symmetry) state to the UUD state (restoring $U(1)$ symmetry).

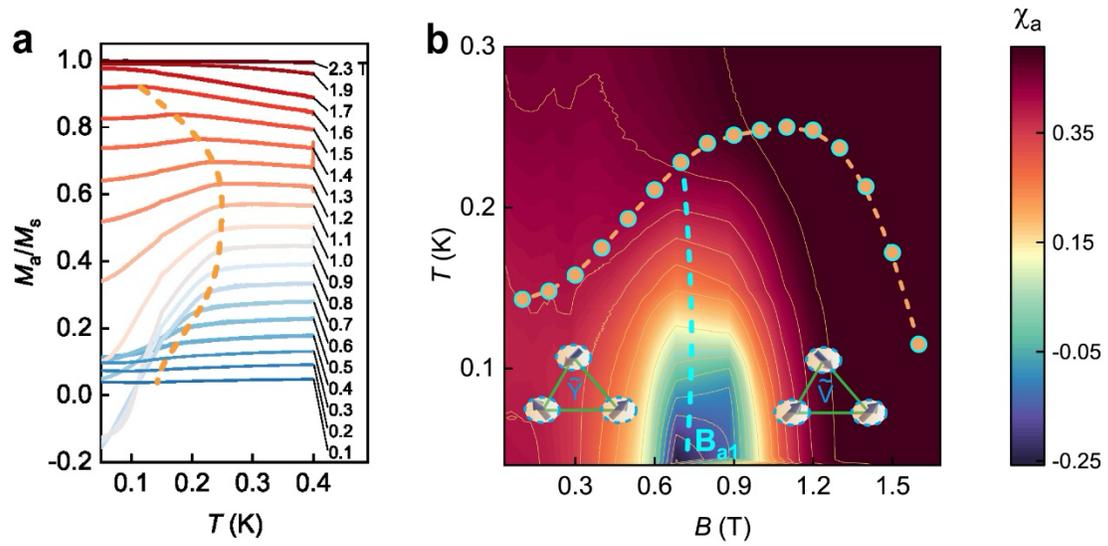

**Fig. 5 *T*-dependent magnetization and the *B-T* phase diagram of NBCP with *B* along the *a*-axis.** (a) Magnetization $M_a$ along the *a*-axis, showing a sharp drop on the left of the orange dashed line with *B* between 0.4 T and 1.5 T, corresponding to the phase transition at $B_{a1}$. (b) The phase diagram shows the possible quantum phase transition marked by the cyan dashed line separating the $\tilde{Y}$ and $\tilde{V}$ states. In the $\tilde{Y}$ state, the spins on the three magnetic sublattices maintain $Z_2$ symmetry, with two spins symmetric about the *a*-direction and one spin exhibiting a positive $S^a$ component. In contrast, the $\tilde{V}$ state breaks this $Z_2$ symmetry, featuring two spins aligned in the same direction and the third spin oriented with an opposite *c* component. The orange dashed line marked the transition from the low-*T* $\tilde{Y}$ and $\tilde{V}$ states to the high-*T* quasi-polarized state.

**Acknowledgment:** We thank Cristian Batista, Junyi Zhang, Yuan Gao, and Wei Li for insightful discussions. The work at the University of Michigan is supported by the Department of Energy under Award No. DE-SC0020184 (magnetization measurements) to Kuan-Wen Chen, Dechen Zhang, Guoxin Zheng, Aaron Chan, Yuan Zhu, Kaila Jenkins, and Lu Li. The work at the University of Tennessee (crystal growth) is supported by the Department of Energy under Award No. DE-SC0020254 to Qing Huang and Haidong Zhou.

**Author contributions:** D.Z. and L.L. conceived the project. D.Z., Y.Z., G.Z., K.C., K.J., and A.C. performed the gradient force magnetometry measurements and data analysis under the supervision of L.L.. Q.H. and H.Z. provided the NBCP single crystals. D.Z., L.Z., and Y.L. prepared the figures. D.Z. wrote the manuscript. All authors discussed the results.

**Competing Interests:** The authors declare no competing financial interests.

**Data Availability:** All data that supports the finding of this work are included as main text and supplementary figures. The data files can be accessed in a depository OSF (Link to be added).

Methods

**Gradient force magnetometry measurement:** Gradient force magnetometry is used to measure the magnetization of NBCP at low temperatures, offering high sensitivity across a wide temperature range. The magnetization measurements were conducted in Oxford Triton200-10 Cryofree Dilution Refrigerator. This technique utilizes thin cantilevers. As shown in Fig. S1a, a typical cantilever consists of a narrow beam with two rectangular pads. Each cantilever is made of beryllium copper foil and has a uniform thickness $T$, typically 10 μm, 25 μm, or 50 μm. As illustrated in the insert of Fig. 2a and Fig. S1b, the setup consists of a cantilever with a metal pad. One pad of the thin cantilever is glued to the top of a sapphire spacer, which is mounted on a sapphire substrate. The NBCP sample is glued on the other metal pad. Both the external magnetic field $B$ and a constant field gradient $\nabla B = 10$ T/m were applied along the $z$-direction. To measure the magnetization along the crystalline $c$-axis and $a$-axis, the sample was glued so that the corresponding crystal axis was aligned with the $z$-direction.

In an external magnetic field, the sample exhibits intrinsic magnetization $M$, generating two torque signals. The dominant signal is the $H$-antisymmetric gradient torque $\tau_{\text{gradient}} = L\,M\,\nabla B$, where $L$ is the distance from the sample center to the cantilever's anchored edge. The second signal $\tau_{\text{torque}} = M \times B$, is a traditional $B$-symmetric torque arising from the non-collinear magnetic field and the sample's anisotropic susceptibility, where $\mu_0$ is the vacuum permeability. These two torque contributions cause the cantilever to deflect, which is detected capacitively.

The measurements are performed as follows. After balancing the capacitance bridge at $C = C_0$, either the magnet field $B$ or temperature $T$ is swept to obtain raw data of capacitance versus $B$ or $T$. The change in capacitance $\Delta C$ corresponds to the torque exerted by the sample on the cantilever metal pad. Since the cantilever and the gold film form a parallel plate capacitor, with $D$ as the distance between the plates and $S$ as the effective cross-sectional area, the capacitance $C$ is given by $C = \varepsilon_0 S/D$, where $\varepsilon_0$ is the vacuum permeability. The angle change of the cantilever $\Delta\theta$ relates to the displacement $\Delta D$ through $\Delta\theta = \Delta D/L$, and the $\tau \propto \Delta\theta$. To infer $\Delta D$ and calculate the torque $\tau$, we ensure that the capacitance change $\Delta C$ remains less than 10% of the zero-field balanced $C_0$. If $\Delta C$ is too large or small, a cantilever with appropriate stiffness (adjusted by its width $W$ and thickness $T$) is selected. To calculate the torque under a magnetic field $B$, we use $\Delta(1/C) = 1/C_0 - 1/C$, with $\tau \propto \Delta(1/C)$, as $\Delta D \propto 1/C_0 - 1/C$. The torque amplitude is calibrated by measuring the sample's saturation magnetization at 2 K.

At relatively low magnetic field $B$, $\tau_{\text{torque}}$ is nearly zero because the sample's crystalline axis is initially aligned with the $z$-direction, resulting in negligible deflection. Consequently, $\tau_{tot}$ is primarily $\tau_{\text{gradient}}$, and the $\tau_{tot} - B$ curve appears nearly centrosymmetric. As $B$ increases, the cantilever deflects slightly, creating a small angle $\theta$ between $B$ and one of the crystalline axes. At higher $B$, the $B$-symmetric torque signal $\tau_{\text{torque}}$ grows quadratically with $B$, and the small angle $\theta$ becomes significant, so the total torque must be considered as $\tau_{tot} = \tau_{\text{gradient}} + \tau_{\text{torque}}$. Since only $\tau_{\text{gradient}}$ is directly proportional to the sample magnetization $M$, we antisymmetrize $\tau_{tot}$ to isolate $\tau_{\text{gradient}}$. This enables direct torque measurement proportional to the sample's magnetization, as shown in Fig. 1E and F. The magnetization is normalized by the saturation magnetization $M_s$ at 50mK. During the field-sweeping measurements, the magnetic field was varied continuously and slowly at a rate of 0.05 T/min at a stable temperature to avoid the magnetocaloric effect.

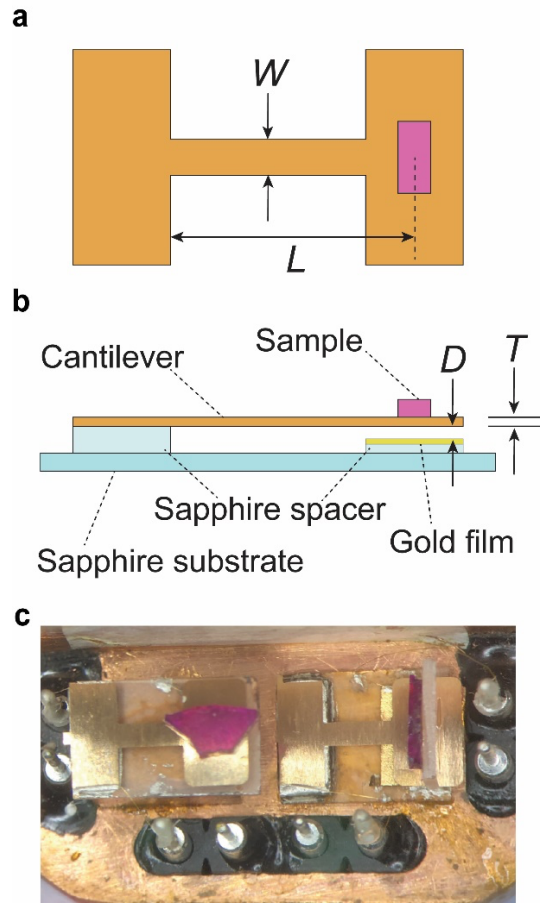

**Fig. S1 Illustration of the gradient force magnetometry setup.** (a) Top view of a typical cantilever. The dimensions of the cantilever depend on the sample magnetization. The typical numbers are: the thickness $T \sim 10$ um, 25 um, or 50 um, the length $L \sim 3 - 4$ mm, and the width $W \sim 0.1 - 0.3$ mm. (b) Schematic of the torque magnetometry setup. A thin cantilever is attached to a sapphire spacer, with the sample mounted at its tip. The cantilever deflection is measured via the capacitance between the cantilever and a gold film on the other sapphire spacer. (c) A photograph of the setup used in the experiments.

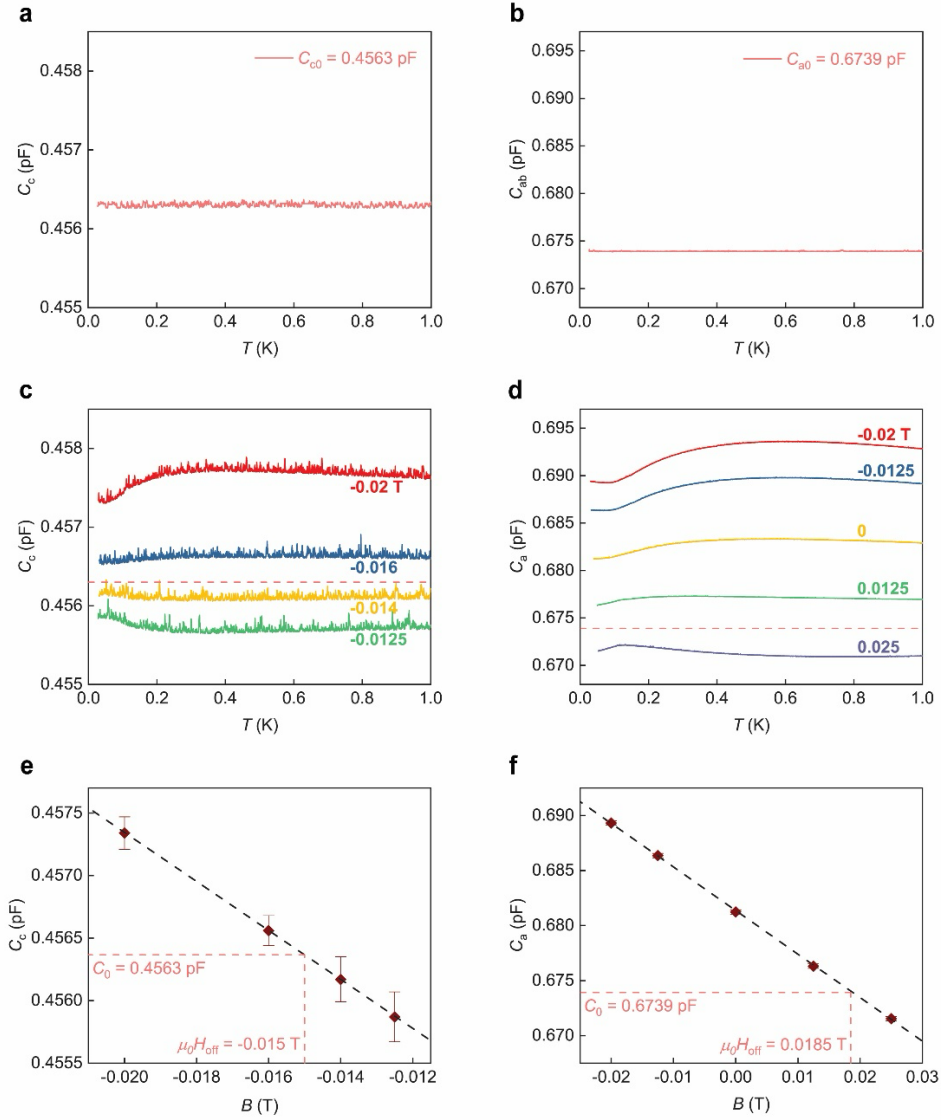

**Fig. S2 Calibrating the center of the magnetic field gradient coil.** Since the sample is measured under a large field gradient of $\nabla B = 10$ T/m, it must be positioned at the center of the gradient coil. Even a small offset of one millimeter from the center can cause a magnetic field deviation of 0.01 T. The sample was first positioned as close to the center as possible. Then, the field offset value for each sample is obtained by doing the calibration. (a-b) The $T$-dependent capacitance readings with $B = 0$ and $\nabla B = 0$, and the crystalline $c$-axis (a) and $a$-axis (b) is along the $z$-direction. (c-d) The $T$-dependent capacitance readings with fixed magnetic fields and the crystalline $c$-axis (c) and $a$-axis (d) along the $z$-direction. The red dashed line marks the capacitance reading under zero field and gradient. The other curves are symmetrical about the red dashed line with a constant offset from zero field. (e-f) Plot and linear fitting of $C_c$ (e) and $C_a$ (f) values at 0.05 K versus field. The field offset value is then found to be $B_{\text{off}} = -0.015$ T for the measurement of $C_c$, and $B_{\text{off}} = 0.0185$ T for the measurement of $C_a$.

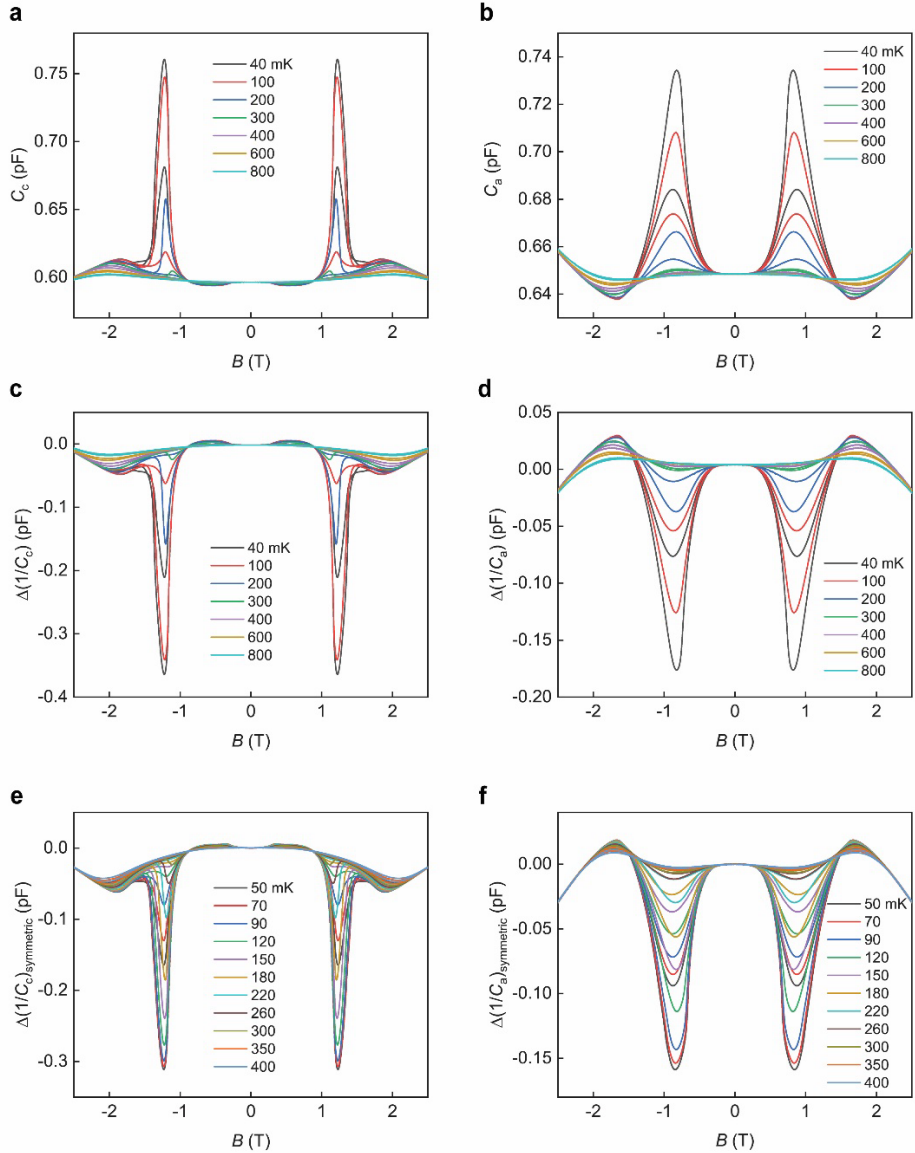

**Fig. S3 Comparison between the symmetric component of $\tau_{\text{tot}}$ under $\nabla B = 10$ T/m and the pure torque signal $\tau_{\text{torque}}$ under $\nabla B = 0$.** (a-b) The $B$-dependent capacitance readings which measure the pure torque signal under $\nabla B = 0$, and the magnetic field is along the crystalline c-axis (a) and a-axis (b). (c-d) The calculated $B$-dependent values of $\Delta(1/C)$ (the torque signal $\tau_{\text{torque}}$) under $\nabla B = 0$, and the magnetic field is along the crystalline c-axis (a) and a-axis (b). (e-f) The $B$-dependent symmetric component of $\tau_{tot}$ under $\nabla B = 10$ T/m, which exhibits a shape and amplitude that closely resembles the torque signal. By doing the antisymmetrization to $\tau_{tot}$, the torque component is canceled to obtain the pure gradient force signal $\tau_{\text{gradient}}$, which is proportional to the sample magnetization. The magnetic field is along the crystalline c-axis (e) and a-axis (f).

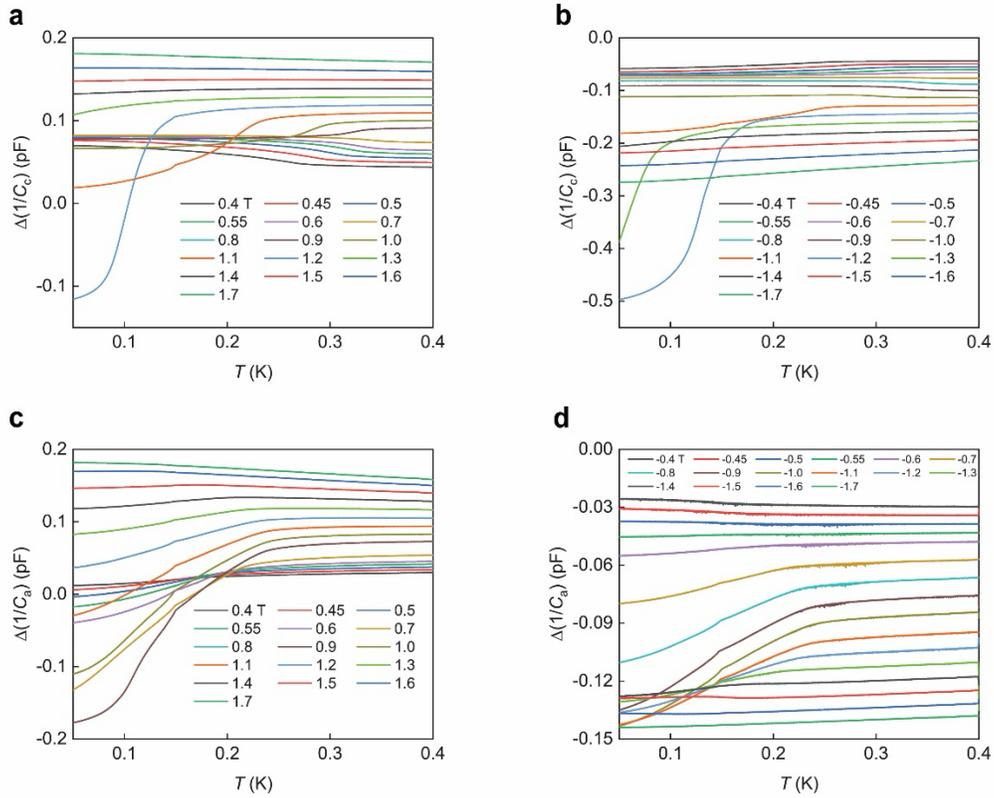

**Fig. S4 The measured temperature dependence of the gradient force signal above 0.415 T at positive and negative fields with the sample mounted on a stiffer cantilever.** The pure gradient force signal proportional to the sample magnetization is obtained by subtracting the positive and negative field data to eliminate the *B*-symmetric torque signal. (a) The temperature dependence of $\Delta(1/C_c)$ under positive field along the crystalline *c*-axis. (b) The temperature dependence of $\Delta(1/C_c)$ under negative field along the crystalline *c*-axis. (c) The temperature dependence of $\Delta(1/C_a)$ under positive field along the crystalline *a*-axis. (d) The temperature dependence of $\Delta(1/C_a)$ under negative field along the crystalline *a*-axis.